\documentclass[superscriptaddress, aps, prl, reprint, twocolumn, amsmath, amssymb, showpacs]{revtex4-1}
\usepackage{graphicx}  % needed for figures
\usepackage{mathptmx}
\usepackage{epstopdf}
\usepackage{dcolumn}   % needed for some tables
\usepackage{bm}        % for math
\usepackage{amssymb}   % for math
\usepackage{amsmath}   % for matrix
\usepackage{amsthm}
\usepackage{chemarrow}
\usepackage{color}
\usepackage{xcolor} 
\usepackage{mathrsfs}
\usepackage{float}
\usepackage{physics}
\usepackage{bbold}
\usepackage{bbm}
\usepackage[toc,page]{appendix}
\usepackage{longtable}
\usepackage{multirow}
\usepackage{upgreek}
\usepackage{epstopdf}
\usepackage{ulem}
\usepackage{graphicx}
\usepackage{lipsum}

\usepackage[]{ezedits}
\defineEdit{WJ}{\color[rgb]{0.5,0.3,0.9}}{\color[rgb]{0.5,0.3,0.9}}
\defineEdit{DB}{\color[rgb]{0,0.3,1}}{\color[rgb]{0,0.3,1}}
\defineEdit{ZTX}{\color[rgb]{1,0.25,0.35}}{\color[rgb]{1,0.25,0.35}}

\usepackage[a4paper, colorlinks=true, linkcolor=blue, citecolor=blue,
pdfauthor={ },
pdftitle={ },
pdfsubject={ },
pdfkeywords={ }]{hyperref}
\usepackage[version=4]{mhchem}
\usepackage{ulem,fancyvrb}

\theoremstyle{plain}
\newtheorem*{theorem*}{Theorem}

\usepackage{titlesec}
\titlespacing*{\section}
{0pt}   % left
{10pt}  % up
{5pt}   % down
\titleformat{\section} % define \section
  {\large\bfseries}   % larger and bold
  {\thesection}       % rank
  {1em}               % distance
  {}                  % after section

\begin{document}

\title{Search for a parity-violating long-range spin-dependent interaction}
 
\date{\today}

    \author{Xing Heng}
	\affiliation{School of Instrumentation Science and Opto-electronics Engineering, Beihang University, Beijing, 100191, China}

\author{Zitong Xu}
    \affiliation{School of Physical and Mathematical Sciences, Nanyang Technological University, Singapore, 639798, Singapore}
	\affiliation{School of Instrumentation Science and Opto-electronics Engineering, Beihang University, Beijing, 100191, China}

\author{Xiaofei Huang}
	\affiliation{School of Instrumentation Science and Opto-electronics Engineering, Beihang University, Beijing, 100191, China}
\author{Dinghui Gong}
	\affiliation{School of Instrumentation Science and Opto-electronics Engineering, Beihang University, Beijing, 100191, China}
\author{Guoqing Tian}
	\affiliation{School of Instrumentation Science and Opto-electronics Engineering, Beihang University, Beijing, 100191, China}	

	\author{Wei Ji}
	\email[Corresponding author: ]{wei.ji@pku.edu.cn}
    \affiliation{School of Physics and State Key Laboratory of Nuclear Physics and Technology, Peking University, Beijing 100871, China}

\author{Jiancheng Fang}
	\affiliation{School of Instrumentation Science and Opto-electronics Engineering, Beihang University, Beijing, 100191, China}	
	    \affiliation{Hefei National Laboratory, Hefei, 230088, China}
        
    \author{Dmitry Budker}
	\affiliation{Johannes Gutenberg University, Mainz 55128, Germany}
	\affiliation{Helmholtz-Institute, GSI Helmholtzzentrum fur Schwerionenforschung, Mainz 55128, Germany}
	\affiliation{Department of Physics, University of California at Berkeley, Berkeley, California 94720-7300, USA}

    \author{Kai Wei}
    \email[Corresponding author: ]{weikai@buaa.edu.cn}
    \affiliation{School of Instrumentation Science and Opto-electronics Engineering, Beihang University, Beijing, 100191, China}
    \affiliation{Hangzhou Extremely Weak Magnetic Field Major Science and Technology Infrastructure Research Institute,
Hangzhou, 310051, China}
     \affiliation{Hefei National Laboratory, Hefei, 230088, China}
     
\begin{abstract} 
High-sensitivity quantum sensors are a promising tool for experimental searches for beyond-Standard-Model interactions. Here, we demonstrate an atomic comagnetometer operating under a resonantly-coupled hybrid spin-resonance (HSR) regime to probe P-odd, T-even interactions. The HSR regime enables robust nuclear-electron spin coupling, enhancing measurement bandwidth and stability without compromising the high sensitivity of spin-exchange relaxation-free magnetometers. To minimize vibration noise from velocity-modulated sources, we implement a multistage vibration isolation system, achieving a vibration noise reduction exceeding 700-fold. We establish new constraints on vector-boson-mediated parity-violating interactions, improving experimental sensitivity by three orders of magnitude compared to previous limits. The new constraints complement existing astrophysical and laboratory studies of potential extensions to the Standard Model.
\end{abstract}

\maketitle

The discovery of parity violation in experiments such as the pioneering study of $\beta$ decay of $^{60}$Co\,\cite{wu1957} and atomic parity violation\,\cite{barkov1978,conti1979}, fundamentally reshaped our understanding of particle physics and contributed to the establishment of the Standard Model. This phenomenon provides a unique opportunity to test the Standard Model through low-energy precision measurements. Ongoing research explores parity violation in atoms, such as cesium\,\cite{wood1997,bouchiat2011atomic}, francium\,\cite{Gwinner2022} and ytterbium\,\cite{tsigutkin2009,antypas_isotopic_2019}. In these systems, parity violation arises from the exchange of $Z$-bosons—the mediators of ``neutral current'' weak interactions—which are distinct from the ``charged current'' interactions mediated by $W^{\pm}$ bosons responsible for the $\beta$ decays as studied in C. S.\,Wu's famous experiment\,\cite{wu1957}. Further exploration of neutral-current parity violation also sheds light on the potential existence of ``new'' $Z'$-bosons, as well as $Z'$-boson-mediated exotic interactions\,\cite{safronova2018search}. 

It is proposed that new exotic spin-dependent forces may exist and that the corresponding interaction potentials may be classified into 16 terms based on their symmetry properties\,\cite{dobrescu2006spin}. These interactions could be mediated by $Z'$ or spin-0 particles such as the axion\,\cite{moody1984new}; some of them could violate parity. Investigation of such interactions could also illuminate the dark matter problem because both $Z'$ and axions are promising dark-matter candidates. The exotic forces are then classified according to their physical coupling constants, providing a unified framework for studying the effects of hypothetical bosons and their interactions\,\cite{fadeev2019revisiting,cong2024spin}. Among these interactions, certain terms dominate in experimental sensitivity, with the axial-vector and vector couplings exhibiting a velocity-dependent parity-violating component\,\cite{cong2024spin}:
\begin{eqnarray}
V_{PV} =  \frac{g_{A}g_{V} {\hbar}}{4\pi}
\left(\hat{\boldsymbol{\sigma}}\cdot\boldsymbol{v} \right)\frac{e^{-{r}/{\lambda}}}r \, , 
\label{eq.v12+13}
\end{eqnarray}
\noindent where \(\hbar\) is the reduced Planck constant, \(\hat{\boldsymbol{\sigma}}\) is the Pauli-matrix vector of the sensing fermion, \(\boldsymbol{v}\) and \(r\) are the relative velocity and distance between the probe fermion and the source fermion, \(\lambda=\hbar/m_b c\) is the force range, $m_b$ is the mass of $Z'$, and $c$ is the speed of light. This potential corresponds to $V_{12+13}$ in\,\cite{dobrescu2006spin,cong2024spin}, with the coupling-strength constant \(g_Ag_V\) related to the coefficient \(f_{12+13}\) as \(f_{12+13}^{ij}=2g_A^ig_V^j\times (1+\frac{m_i} {m_j})\), where \(i,j\) label various fermion pairs (e.g., \(e-N\), \(n-N\), \(p-N\)). Current astrophysical observations exhibit gaps in the constraints for spin-dependent interactions, motivating direct tests via tabletop experiments.\,\cite{cong2024spin,wei2023ultrasensitive}. 

In this work, we utilize two lead blocks with a high mass density as the source and employ a state-of-the-art spin-exchange relaxation free (SERF) comagnetometer with an ultrahigh energy resolution\,\cite{wei2023ultrasensitive,heng2023ultrasensitive,xu2023darkmatter} as the sensor to search for exotic particles and velocity-dependent parity-violating interactions. By operating the SERF comagnetometer in the resonantly-coupled hybrid spin-resonance (HSR) regime, we significantly enhance the measurement bandwidth, leading to improved stability while preserving the high sensitivity of the SERF magnetometer. Vibrations induced by the rotating masses is a noise source for these measurements, which we mitigate with a multistage scheme that combines a vibration-isolated foundation with a vacuum system enclosing the sensor. This provides a more than 700-fold suppression of the vibration noise. We establish the most stringent experimental constraints over a force range of 0.03 to 400 meters, particularly improving upon previous limits\,\cite{su2021search,yan2015searching} by three orders of magnitude at \(\lambda=5\)\,m.

\section*{Results}

\textbf{Principle} 

\begin{figure}[htb]
    \centering
\includegraphics[width=0.99\linewidth]{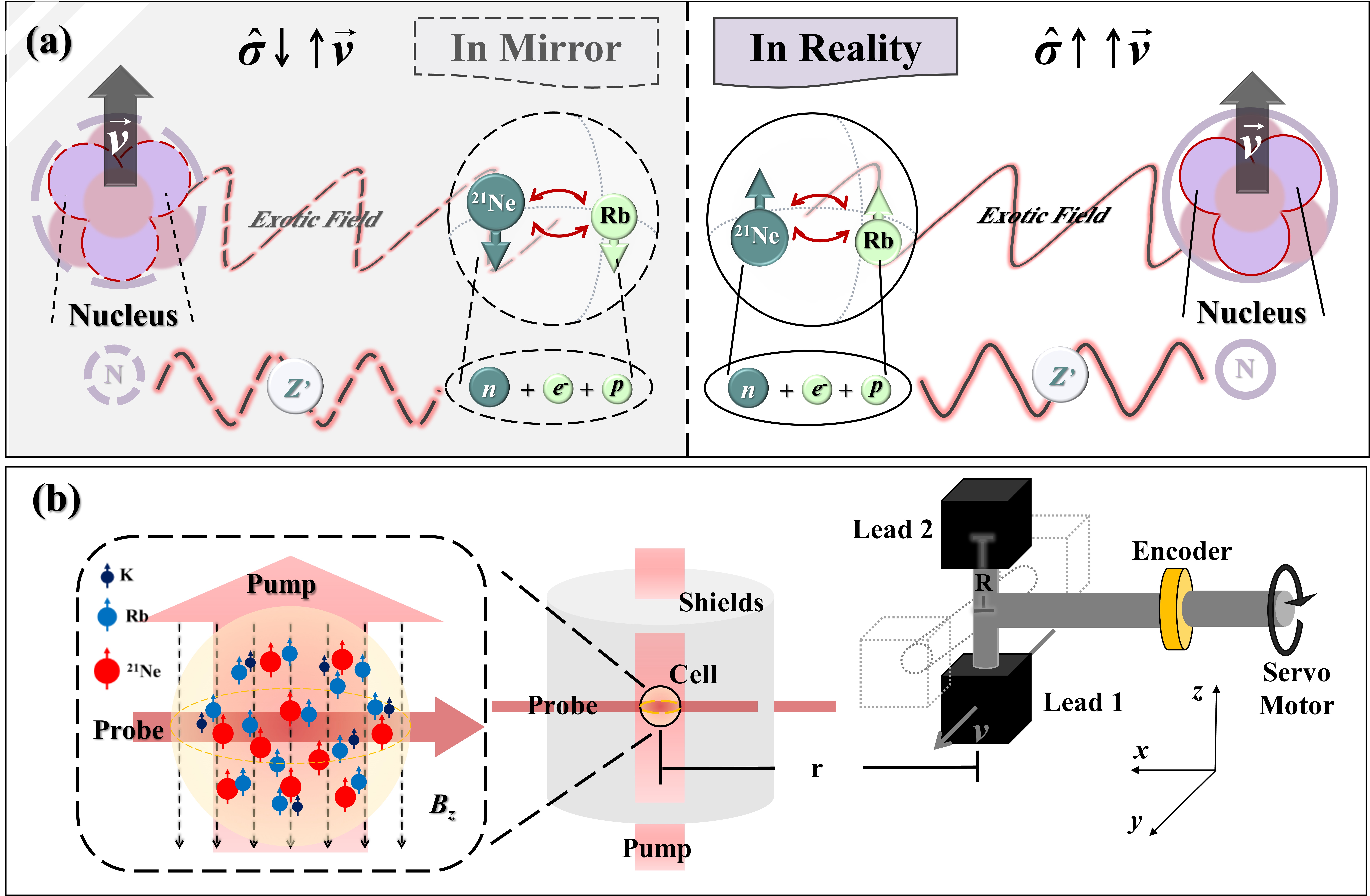}
\caption{Principle of the experiment. (a) Parity can be tested by the exotic spin- and velocity-dependent interaction \(V_{12+13}\). Notably, the induced pseudomagnetic field exhibits a sign change upon spatial inversion, highlighting the inherent asymmetries explored in this investigation. (b) Experimental setup. The servo motor drives two cubic lead blocks at a rotation frequency of 3\,Hz, inducing a pseudomagnetic field sensed with an ensemble of polarized hybrid spins. When a lead block reaches the lowest position, the center of the block is in the horizontal plane of the atomic reservoir (vapor cell), at which point the centers of the lead block and the vapor cell are separated by a distance \(r\)=52.5\,cm. }
\label{Fig.setup} 
\end{figure}

Parity violation by the exotic force is illustrated in Fig.\,\ref{Fig.setup} (a). In the mirror-reflected framework, the spins exhibits inverted polarization compared to the physical reality, while the velocity does not, which leaves them parallel in reality and antiparallel in the mirror. Therefore, the product of velocity and spin changes sign by mirror reflection (and the parity operation), which indicates the parity non-conservation of Eq.\,(\ref{eq.v12+13}).

\begin{figure}[htb]
    \centering
\includegraphics[width= 0.95 \linewidth]{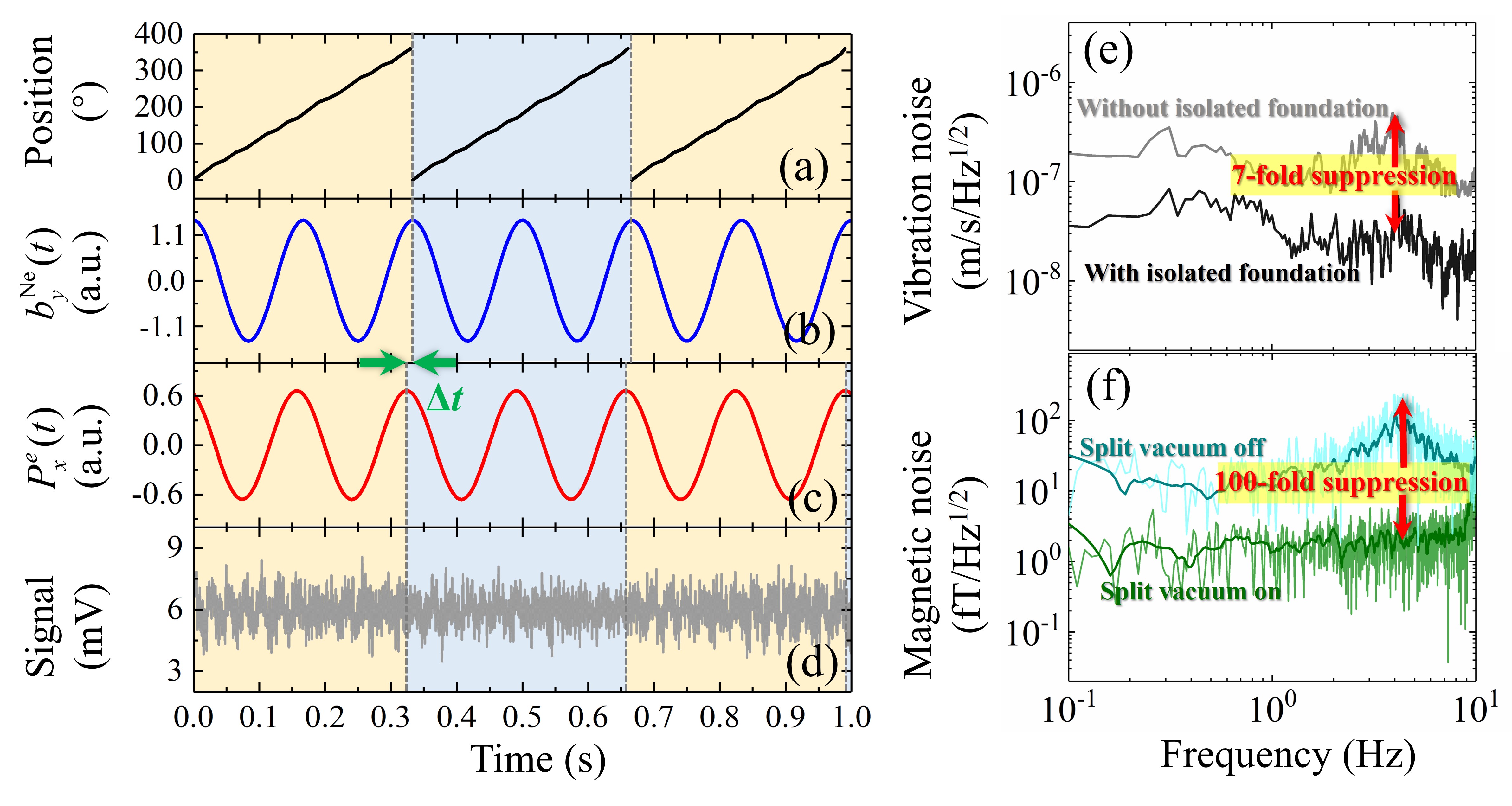}
\caption{Data acquisition and simulation process. (a) The optical encoder measures the rotation angle (0\,\(^\circ\) to 360\,$^\circ$) in real-time to obtain the position of Lead 1. An angle of 0\,$^\circ$ indicates that Lead 1 is at the lowest point. (b) As the motor rotates, the pseudomagnetic field \(b_y^{\text{Ne}}\) of double frequency induced by the two lead blocks is simulated. (c) The simulated response of the co-magnetometer to \(b_y^{\text{Ne}}\) displays a delay $\Delta t$, which corresponds to the phase relative to the \(b_y^{\text{Ne}}\) input. (d) The measured response of the co-magnetometer. Vibration noise in the experiment is observed and multistage vibration suppression is conducted. (e) Comparison of vibration noise with (black line) and without (gray line) implementing the first-level vibration isolation foundation. Following the installation of the isolation foundation, the inherent vibration peaks of the ground in the frequency range of interest have been suppressed by a factor of 7. (f) Assessment of the second-level vibration suppression is conducted by measuring the equivalent magnetic noise of the comagnetometer signal with the split vacuum system on (olive line) and off (cyan line). The specialized split vacuum system, operating at a pressure below 50\,mBar, achieves a noise suppression exceeding 100 times. The overall multistage vibration suppression factor exceeds 700.}
\label{Fig.dataprocess} 
\end{figure}

The experimental setup is depicted in Fig.\,\ref{Fig.setup} (b). Two cubic lead blocks, each with a side length of 10.00\,cm, are mounted on the axis of a high-power servomotor. The lead blocks have a high mass density of \(11.3\,\text{g/cm}^3\), resulting in a correspondingly high nucleon density (\(6.8 \times 10^{24}\,\text{/cm}^3\))\,\cite{cong2024spin}, which makes them an ideal choice of mass source\,\cite{lee2018improved,liang2023new,crescini2022search}. The servomotor rotates the two lead blocks at approximately 3\,Hz, corresponding to the exotic force modulation at 6\,Hz due to symmetry. The centers of the lead blocks rotate in a circle with a radius of 50.0\,cm, with the bottom plane of the blocks aligned with the atomic vapor at the lowest point of rotation. An optical encoder is used to monitor the rotational angle in real time with a precision of \(\pm4.9\)\,\(\mu\)rad, and one set of data is shown in Fig.\,\ref{Fig.dataprocess} (a).

The K–Rb–\(^{21}\)Ne comagnetometer used in this experiment is similar to that described in Refs.\,\cite{xu2023darkmatter,heng2023ultrasensitive}. A 12-mm-diameter spherical cell, containing 2280\,torr of \(^{21}\)Ne, 70\,torr of N\(_2\), and K/Rb vapor (density ratio \(\sim\)1:100 at 195\,\(^{\circ}\)C), is enclosed within a five-layer magnetic shield \(\mu\)-metal and Mn-Zn ferrite). Hybrid optical pumping along the $z$ axis enhances the polarization uniformity of the alkali spin and the hyperpolarization efficiency of the noble gas spins: a circularly polarized resonant laser directly polarizes the K atoms, while spin-exchange collisions transfer polarization to Rb atoms and \(^{21}\)Ne nuclei. The exotic force can couple to the electron, proton and neutron spins in the alkali atoms. Here we take the \(^{21}\)Ne nuclei as an example. The precession of \(^{21}\)Ne nuclei, induced by the parity-violating pseudomagnetic field \(b_y^{\text{Ne}}\), generates a real magnetic field that can be detected by Rb atoms. The resulting dynamics of the Rb atoms are measured via the optical rotation of off-resonant polarized light propagating along the \textit{x}-direction. The exotic field \(b_y^{\text{Ne}}\) generated by the source mass can be inferred from the measured optical rotation signal with a conversion factor \(K_{b_y^n}\), which is detailed in Eq.\,(\ref{eq.HSRresponse}).

\textbf{Hybrid spin-resonance regime}

In long-term precision measurements, the drift of system parameters and variations in environmental conditions can significantly affect the accuracy and stability of measurement results. This necessitates a careful balance between high sensitivity and long-term stability in the measurement of weak pseudomagnetic fields. Operating the coupled atomic ensemble in the resonantly-coupled HSR regime is particularly suitable for this purpose. By applying a static magnetic field \(B_z\) along the \textit{z}-axis (the spin-polarization axis) with a magnitude equals to the magnitude of the effective magnetic field from the Fermi-contact interactions between the Rb and $^{21}$Ne atoms\,\cite{xu2022critical}, the Larmor precession of the \(^{21}\)Ne spins and the Rb atoms becomes strongly coupled. In this scenario, the relaxation of noble-gas spins is influenced by alkali spins, resulting in a broader bandwidth of nuclear spins, and the regime exhibits superior stability across a broad frequency range (details can be found in METHODS). The response of the comagnetometer to the oscillating exotic field coupled to noble-gas spins, represented as \(b^{\text{Ne}}_y(t) = b^{\text{Ne}}_{y0}\text{cos}(\omega t)\), can be expressed as:
\begin{eqnarray}
P^e_x(t) = K_{b_y^n}b^{\text{Ne}}_{y0}\text{cos}(\omega t+\phi_{b_y^{\text{Ne}}}) \, ,
\label{eq.HSRresponse}
\end{eqnarray}
\noindent where \(K_{b_y^n}\) relates the pseudomagnetic field with the spin polarization along \textit{x}-axis. This factor depend on the longitudinal spin polarizations \(P^{e/n}_z\), the transverse relaxation rates of both spins and the effective magnetic fields from Fermi-contact interactions (see details in\,\cite{xu2022critical}), and \(\phi_{b_y^{\text{Ne}}}\) presents the phase shift  of the optical rotation signal due to  \(b_y^{\text{Ne}}\).
In experimental setups, classical magnetic fields are typically employed to calibrate the response factor of the atoms to the exotic field \(K_{b_y^n}\). This methodology is commonly utilized in experiments seeking new physics beyond the Standard Model\,\cite{padniuk2024universal,xu2023darkmatter}.

The HSR regime of the comagnetometer demonstrates a bandwidth up to 25\,Hz, shown in Fig.\,\ref{Fig.stability}(a), which tends to cover multiple harmonic components of \(b_y^{\text{Ne}}\). By changing the rotation frequency of the lead blocks, exotic signals are in the frequency region where the noise performance of the HSR comagnetometer is optimal. Utilizing the methodology outlined in\,\cite{wei2022constraints}, we simulated the pseudomagnetic signal \(b_y^{\text{Ne}}\), with the primary input parameters for the simulation detailed in Table ~\ref{tab:errors}. The simulated response is illustrated in Fig.\,\ref{Fig.dataprocess} (b). The modulated exotic field generated by the source mass can be detected by the comagnetometer as \(b_y^{\text{Ne}}=\zeta^{n,p}/\mu_{\text{Ne}} \int{V_{PV} dV}\), where \(\zeta^n=0.58\) and \(\zeta^p=0.04\) are the fraction factors for neutron and proton spin polarization in the \(^{21}\)Ne nucleus \cite{wei2022constraints, almasi2020new}, while \(\mu_{\text{Ne}}\) stands for the magnetic moment of the \(^{21}\)Ne nucleus. Figure\,\ref{Fig.dataprocess} (c) presents the simulated $P^e_x(t)$, derived from Eq.\,\ref{eq.HSRresponse}, employing the calibrated parameters of the K-Rb-\(^{21}\)Ne ensemble in METHODS. We further investigate the phase shifts between the HSR response and the exotic field \(b_y^{\text{Ne}}\). At a modulation frequency of 6\,Hz (corresponding to rotation frequency of 3\,Hz), the phase shift \(\phi_{b_y^{\text{Ne}}} = 10.1 \pm 5.6\,^\circ \) corresponds to a time delay  \(\Delta t = 4.7 \pm 2.6 \, \text{ms}\), which aligns with the respective signal segments illustrated in Fig.\,\ref{Fig.dataprocess} (b) and (c). Figure\,\ref{Fig.dataprocess} (d) displays the corresponding response of the comagnetometer, where fluctuations in the experimental data are attributed to resonance vibrations of the mechanical structure of the setup. The vibration noise, a primary source of uncertainty in the experiment, is mitigated by an overall factor of 700 (shown in Fig.\,\ref{Fig.dataprocess} (e) and (f)) through a multistage isolation approach.  

\begin{table}[htb]
\centering
\caption{Summary of calibrated parameters and systematic errors. The corrections to \(g_A^ng_V^N\) for \(\lambda=5\)\,m are listed.}
\begin{tabular}{ccc}
    \hline
    \hline
    Parameter & Value  &  \(\Delta g_A^ng_V^N (\times 10^{-39})\) \\ 
    \hline
    Mass of Lead \(M\) (kg) &\(12.00 \pm 0.01\) &\(<0.01\)\\ 
    \hline
    Position of \(X\) (cm) &\(52.5 \pm 5.5\) &\(0.28\)\\ 
    \hline
    Position of \(Y\) (cm) &\(0.0 \pm 1.0\) &\(<0.05\)\\ 
    \hline
    Position of \(Z\) (cm) &\(0.0\pm 1.0\) &\(<0.05\)\\     
    \hline
    Modulation Frequency \(f_m\) (Hz)  &\(6.00 \pm 0.14\) &\(0.06\)\\ 
    \hline
    Ring radius \(R\) (cm) & \(50.0 \pm 1.0\)  &\(0.06\)\\
    \hline
    Calibrated \(K_{b_y^n}\) ($\mu$V/fT) & \( 0.193 \pm0.016\) & 0.77 \\
    \hline
    \multirow{2}{*}{Phase uncertainty \(\phi_{b_y^{\text{Ne}}}\) (\(^{\circ}\))} & \multirow{2}{*}{\(10.1 \pm 5.6\)} & +1.4 \\ 
        & & -2.3 \\
    \hline
   Vibration noise (m/s/Hz\(^{1/2}\)) & \(<5.6 \times 10^{-10}\) & \(<5.4\) \\
    \hline
    Final \(g_A^ng_V^N (\times 10^{-39})\) &\(5.31\) & 5.88(syst)\\ 
    \hline
    (\(\lambda=5\)\,m) &  & 12.38(stat) \\ 
    \hline
    \hline
\end{tabular}
\label{tab:errors}
\end{table}
    
\textbf{New constraints on parity-odd interactions} 

\begin{figure}[htb]
    \centering
\includegraphics[width= 0.99 \linewidth]{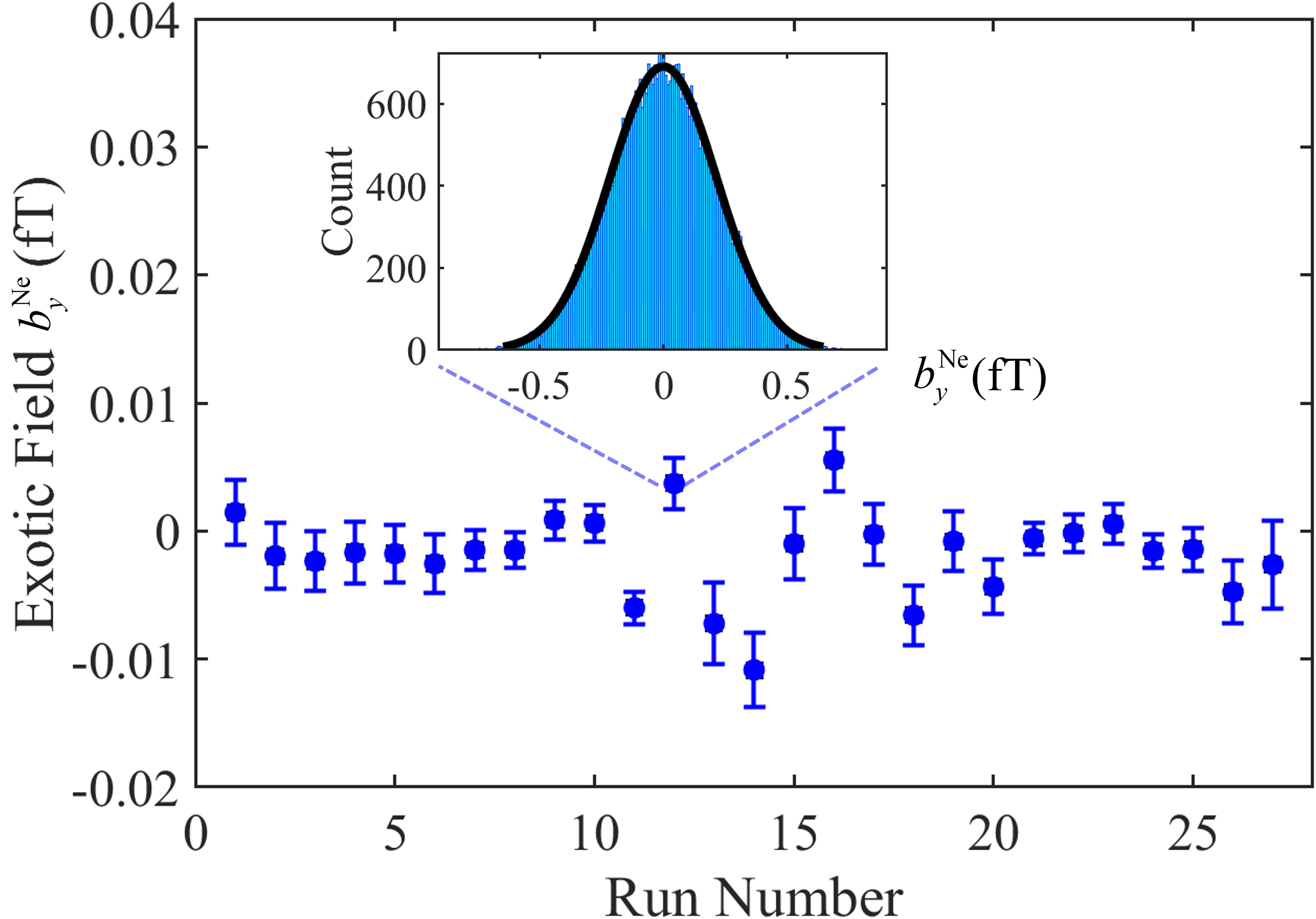}
\caption{ Experimental results of the exotic field \(b_y^{\text{Ne}}\). Each point represents an average of about one 4-h dataset. The error bars represent the statistical of the comagnetometer combined in quadrature. The distribution of \(b_y^{\text{Ne}}\) for one data set is shown in the insert, with the black curve being a Gaussian fit with reduced \(\chi^2\) =1.16. The exotic field \(b_y^{\text{Ne}}\) is measured to be \((1.8 \pm 4.2_\text{stat})\)\,aT. }
\label{Fig.experiment} 
\end{figure}

The total duration for collecting the pseudomagnetic field is 108\,h, during which the measurements are conducted in time series of 4\,h intervals to ensure calibration stability and experimental convenience\,\cite{wei2022constraints}.   The amplitude of the dominant harmonic components of the pseudomagnetic field is analyzed with a weighting method (please refer to\,\cite{ji2018new,xu2025new,ji2023constraints}) to mitigate potential slow drifts within the system, and weighted processing is performed according to the Fourier coefficients \(c_k\) of each harmonic component (see in METHODS). The measured \(b_y^{\text{Ne}}\) is found to be \((1.8 \pm 4.2_\text{stat}\pm 2.0_\text{syst})\)\,aT, as depicted in Fig.\,\ref{Fig.experiment}. The results account for various system uncertainties presented in Table\,\ref{tab:errors}, including statistical uncertainty and systematical errors related to source mass, position, rotation radius, modulation frequency, calibration factor, and signal phase. The primary systematic error arises from vibrations, especially the acoustic coupling between source-mass rotation and output signals. Measurements using a seismometer indicates that the mass source system induced apparatus vibration is approximately \(3.9 \times 10^{-7}\)\,m/s/Hz\(^{1/2}\). To mitigate this major systematic error in measurements, we design a specialized split vacuum chamber and install the HSR comagnetometer on an isolated foundation\,\cite{liu2023modeling,tian2025active}. Multistage isolation effectively suppresses vibration noise by over a factor of 700, achieving a level below \(5.6 \times 10^{-10}\)\,m/s/Hz\(^{1/2}\).

\begin{figure}[htb]
    \centering
\includegraphics[width= 0.99 \linewidth]{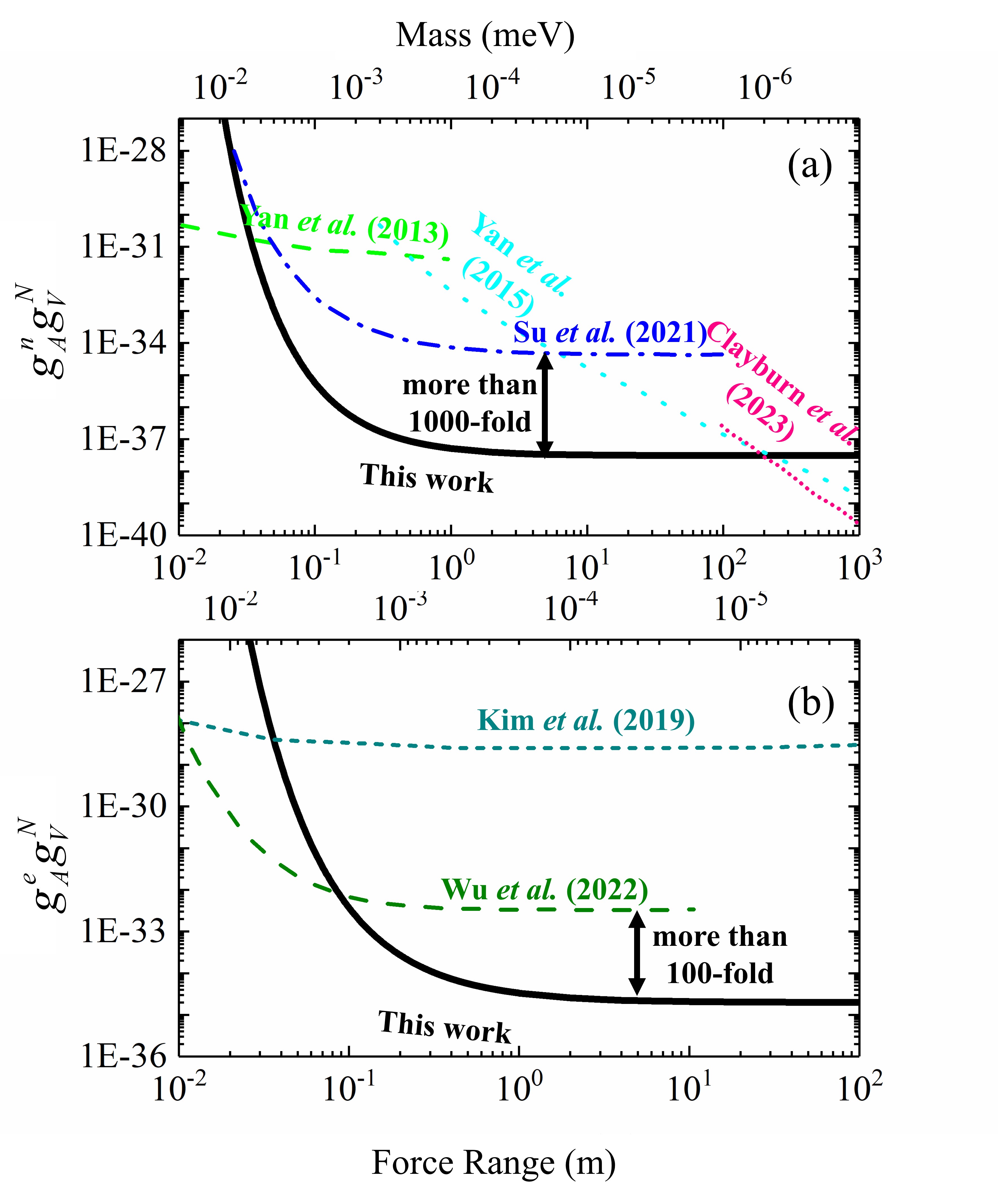}
\caption{The experimental limits on coupling coefficients. (a) Constraints on the coupling constants \(g_A^ng_V^N\) (\(1\sigma\)) as a function of interaction force range. The black solid line is the current constraints. The green dashed line, “Yan \textit{et al.} (2013)” is from Ref.\,\cite{yan2013new} that search for parity violation in neutron spin rotation in liquid \(^{4}\)He; the cyan dotted line, “Yan \textit{et al.} (2015)” is from Ref.\,\cite{yan2015searching} that considers the Earth as an unpolarized source; and the “Clayburn \textit{et al.} (2023)” from Ref.\,\cite{clayburn2023using}, the pink dotted line, uses Earth to search for long-range spin-velocity interactions; the blue dashed-dotted line, “Su \textit{et al.} (2021)” is from Refs.\,\cite{su2021search} that uses a spin-based amplifier. (b) Constraints on the coupling constants \(g_A^eg_V^N\) (\(1\sigma\)) as a function of interaction range. The black solid line is the current constraints. The dark-cyan dashed line, “Kim \textit{et al.} (2019)” is from Ref.\,\cite{kim2019experimentalNC} that take the fraction of electron spin into account; the olive dashed line, with modification as well, “Wu \textit{et al.} (2022)” is from Refs.\,\cite{wu2022experimentalPRL} that uses two BGO masses and an atomic magnetometer array.}
\label{Fig.result} 
\end{figure}

Figure\,\ref{Fig.result} illustrates the new limits on the dimensionless coupling coefficient established by the experiment. Considering the contribution of neutron spins in \(^{21}\)Ne, we established the strongest constraints on the interaction between neutrons and nucleons, as depicted by the solid black line in Fig.\,\ref{Fig.result} (a). At a force range of \(\lambda=5\)\,m, the coupling constant is \(g_A^ng_V^N=(5.3\pm12.4_{\text{stat}}\pm5.9_{\text{syst}}) \times 10^{-39}\), with \(95\%\) confidence level of \(|g_A^ng_V^N| \leq 2.9 \times 10^{-38}\), signifying a three-orders-of-magnitude improvement over previous constraints\,\cite{su2021search}. Our findings set the currently most stringent limits on parity-odd interactions within a force range of 0.03 to 400\,m.
Additionally, the exotic force coupling to electron spins in the comagnetometer can be deduced with the same method as the $^{21}$Ne. We need to replace the magnetic moment of $^{21}$Ne to Rb, and use the response curve of the Rb atoms to the exotic field. For Rb atoms with \(50\%\) polarization, the proton's fraction of spin is approximately 0.29, while the electron's is about 0.13\,\cite{ji2023constraints}. We present the result for \(e-N\) in Fig.\,\ref{Fig.result} (b). The corresponding result for \(p-N\) can be obtained by rescaling the line of \(e-N\) according to the ratio of their fraction of spin. We obtained \(b^e_y\)=(\(0.078\pm0.183_{\text{stat}} \pm0.171_{\text{syst}}\))\,aT, corresponding to a coupling constant \(g_A^eg_V^N=(2.8\pm6.7_{\text{stat}}\pm6.2_{\text{syst}}) \times 10^{-36}\) at \(\lambda=5\)\,m, establishing a limit at the \(95\%\) confidence level of \(|g_A^eg_V^N| \leq 1.9 \times 10^{-35}\), representing an enhancement of more than two orders of magnitude over the previous limit established in Ref.\,\cite{wu2022experimentalPRL}. Please note that the Ref.\,\cite{wu2022experimentalPRL} and Ref.\,\cite{kim2019experimentalNC} didn't take the fraction of spin into account, and we rescaled their result in the plot with fraction of spin accordingly. 

\section*{Discussion}

In contrast to previous experiments relying on self-compensation (SC)\,\cite{lee2018improved} or nuclear magnetic resonance (NMR)\,\cite{su2021search} modes for exotic field detection, the HSR regime enhances measurement stability by broadening the sensor bandwidth while retaining the exceptional sensitivity of SERF comagnetometers. Demonstrated results show a 45\,dB improvement in disturbance rejection (see Fig.\,\ref{Fig.stability}(b) for details). Unlike the SC mode that suppresses low-frequency magnetic noise by balancing external and effective magnetic fields (\(B_z = -B_z^e-B_z^n\)), the HSR regime operates under the condition \(B_z\approx-B_z^n\)\,\cite{xu2023darkmatter}, enabling synchronized dynamics between alkali and noble-gas spin ensembles. The magnetic suppression factor, defined as the ratio of responses to a classic magnetic field \(B_y\) and a pseudomagnetic field \(b^{\text{Ne}}_y\)\,\cite{wei2023ultrasensitive}, reveals five-fold suppression for slowly varying magnetic fields below 40\,mHz. This suppression capability is indispensable for long-term stability, as it reduces the parameter drifts induced by environmental perturbations (e.g., temperature variations and light-intensity fluctuations) over extended measurement durations. At the same time, the HSR regime inherits the signal amplification advantage of NMR, where spin-ensemble coupling induces an approximately 100-fold enhancement of the effective magnetic field sensed by alkali spins within the HSR frequency range. By simultaneously extending the bandwidth and improving the signal-to-noise ratio (SNR), the HSR regime offers a robust technique for precision measurements.

A major challenge in this experiment arises from vibration noise induced by mass motion. The multistage isolation, integrating the vibration isolation foundation with split vacuum system, totally achieves 700-fold suppression of vibration noise, sheding light on quantum technologies requiring sub-picometer stability such as gravitational wave interferometry\,\cite{aasi2015advanced}. Future research will optimize pseudomagnetic field detection through enhanced spin-coherence materials and multilayer magnetic shielding architectures. Precision metrology incorporating quantum control protocols and machine learning-assisted noise suppression\,\cite{duan2025concurrent} will minimize systematic uncertainties. Implementation of non-classical spin states\,\cite{troullinou2021squeezed} promises to surpass standard quantum limits, thereby extending applications in fundamental interaction studies.

In this work, we demonstrate a resonantly-coupled hybrid spin-resonance regime in atomic ensembles to probe parity-violating interactions, improving constraints on the axial-vector coupling \(g_Ag_V\) by three orders of magnitude. This advance establishes the HSR method as a powerful tool for precision tests of fundamental symmetries. Future refinements may extend its sensitivity to beyond-Standard-Model physics and other exotic spin-dependent interactions. The presented experiment continues the seven-decade tradition of atomic parity violation studies that started with the visionary proposal of Ya.\.B.\,Zel'dovich\,\cite{zel1959parity} and is augmented today with the searches for long-range exotic parity-violating forces with ultrasensitive atomic magnetometers, as presented here.

\section*{Methods}
\begin{figure}[htb]
    \centering
\includegraphics[width= 0.95 \linewidth]{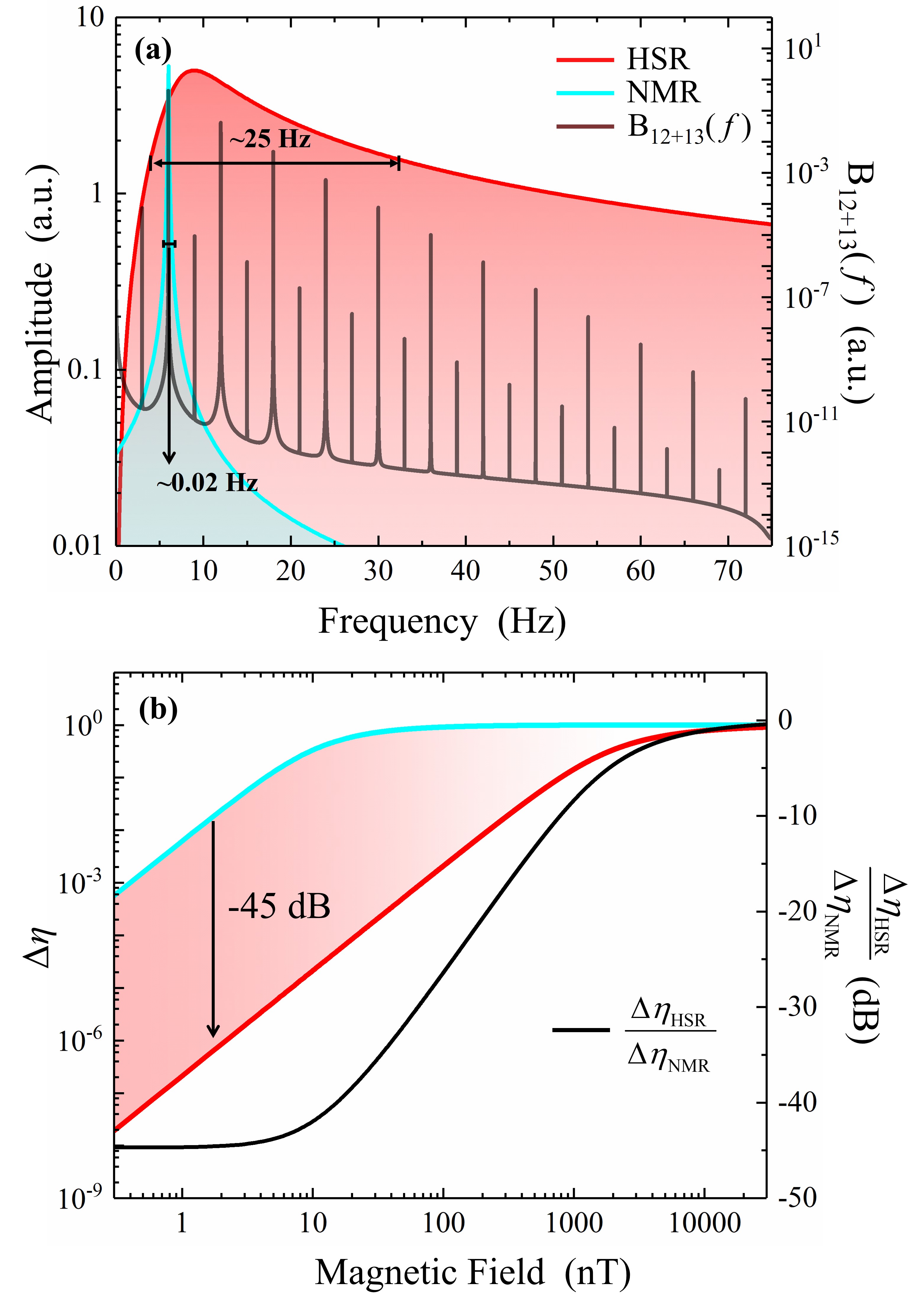}
\caption{Effective bandwidth and disturbance rejection of K-Rb-$^{21}$Ne comagnetometers operating in different regimes. (a) Comparison of bandwidths in HSR (red line) and NMR (cyan line) regimes. The bandwidth of HSR is more than three orders of magnitude greater than that of NMR, which can effectively cover multiple harmonic components of the pseudomagnetic signal \(B_{12+13}\) (gray line). (b) Comparison of disturbance rejection in the HSR and NMR regimes. The relative change in output amplitude of the system under magnetic fluctuations can be quantitatively analyzed using the rejection ratio \(\Delta\eta\) in (\ref{eq.rejection ratio}). \(\Delta\eta\) close to zero indicates strong disturbance rejection. The colored gradient areas illustrate the advantages of HSR over NMR in terms of disturbance rejection. Within the fluctuation range of 0.3 to 10 nT (for \(^{3}\)He\,\cite{lee2018improved} ranges from 0.03 to 1 nT, while 0.09 to 3 nT for \(^{129}\)Xe\,\cite{su2021search}), the disturbance rejection of HSR is enhanced by 45 dB compared to NMR.}
\label{Fig.stability} 
\end{figure} 

\textbf{Stability enhancement} 

In the resonantly-coupled HSR regime, the comagnetometer maintains its ultrahigh sensitivity while achieving a bandwidth on the order of tens of hertz. Compared to NMR magnetometers, the achieved high bandwidth can cover the multiple harmonic components of the pseudomagnetic field, as illustrated in Fig.\,\ref{Fig.stability}(a).

Comagnetometers operating in the HSR regime demonstrate superior disturbance rejection capabilities across a wide frequency range. To quantitatively assess the disturbance rejection performance of the magnetometer, we utilize the rejection ratio, a metric commonly employed in the field of control systems. The rejection ratio is defined as the ratio of the disturbance signal relative to that of the reference input signal. It can be mathematically expressed as:
\begin{eqnarray}
\Delta\eta = \left | \frac{A(f_0)-A(f_0+\Delta f)} {A(f_0)} \right | \, , 
\label{eq.rejection ratio}
\end{eqnarray}
\noindent where \(A(f_0)\) represents the response of the magnetometer to the resonant frequency \(f_0\) and \(\Delta f\) indicates the fluctuations in \(f_0\) caused by variations in experimental parameters such as atomic density, magnetic field, and light intensity. When the K-Rb-\(^{21}\)Ne magnetometer operates in the HSR regime, the response \(A_{\text{HSR}}\propto fQ/\sqrt{(R^e_2/f)^2+[(fQ)^2-\gamma_e \gamma_n B_z^{e} B_z^{n}]^2}\)\,\cite{xu2022critical,xu2023darkmatter}, where \(\gamma_{e/n}\) denote the gyromagnetic ratios for electrons and nuclei, respectively, while \(B_z^{e/n}\) represent the effective magnetic fields on electronic and nuclear spins arising from Fermi-contact interactions between them, and \(Q\) is the slowing-down factor for alkali atoms \cite{wei2023ultrasensitive,heng2023ultrasensitive}. In contrast, for the NMR regime, \(A_{\text{NMR}}\propto 1/\sqrt{(R^n_2)^2+(f-\gamma_n B_z^{\text{eff}})^2}\)\,\cite{su2021search,xu2024constraining},  where \(R_2^{e/n}\) are the transverse relaxation rates of the alkali spin and the noble-gas spin, and \(B_z^{\text{eff}}\) denotes the total effective magnetic field detected by the \(^{21}\)Ne nuclei. The rejection ratios for the two regimes are illustrated in Fig.\,\ref{Fig.stability}(b) with calibrated parameters of the K-Rb-\(^{21}\)Ne ensemble: \(R_2^n \approx \text{0.005\,s}^{-1},~R_2^e \approx \text{3900\,s}^{-1},~Q \approx 7.6,~ B_z^{e} \approx 83\,\text{nT}\), \(B_z^{n} \approx 468\,\text{nT}\), and \(B_z^{\text{eff}} \approx 1.8\,\text{pT}\). 

As shown in Fig.\ref{Fig.stability}(b), the disturbance rejection of both regimes degrades significantly when magnetic deviations exceed the response bandwidths. The HSR regime, with a bandwidth of tens of hertz ($\sim$ 10\,µT), outperforms the NMR regime, which operates at millihertz bandwidths ($\sim$10\,nT). The broader bandwidth in HSR enables effective suppression of frequency fluctuations caused by the environmental parameters such as the residual magnetic fields and non-constant temperature, enhancing pseudomagnetic field measurement stability. For weak pseudomagnetic fields, even minor frequency drifts can severely degrade fidelity, potentially rendering prolonged measurements unreliable. HSR not only improves disturbance rejection but also maintains signal integrity over extended durations. Within a 0.3$\sim$10\,nT fluctuation range, HSR achieves a 45\,dB improvement over NMR, highlighting its potential for high-precision applications requiring robust stability against environmental noise.

\textbf{Data processing} 

The data processing methodology is consistent with the approach delineated in Ref.\,\cite{wu2022experimentalPRL,wu2022searching}. The equivalent pseudomagnetic field comprises multiple harmonic components with a fundamental frequency denoted by \(f_m\), as illustrated in Fig.\,\ref{Fig.stability} (a), and the mathematical representation of the measured signal can be expressed as \(b(t) = g_Ag_V \sum_k [c_k \text{cos}(2\pi k f_m t+\phi)]  +n(t)\), where \(c_k\) denotes the Fourier coefficients of the various harmonic components and can be obtained through numerical integration. The term \(\phi\) corresponds to the initial phase factor of the system, derived from the phase signal of the encoder, while  \(n(t)\) accounts for noise in the measurement. Within the bandwidth, the interaction coupling constant \(g_Ag_V\) can ultimately be determined as \( \overline{g_Ag_V} = \sum_k(c_k^2 g_Ag_V|_k)/\sum_k(c_k^2)\) (similar in \cite{wu2022searching}). Here, the \textit{k}-th harmonic interaction coupling constant is defined by 
\begin{eqnarray}
g_Ag_V|_k = \frac{2f_m}{c_k M} \times\int_0^{M/f_m} \text{cos}(2\pi k f_m t+\phi)b(t) \, dt \, , 
\label{eq.data process1}
\end{eqnarray}
with \(M/f_m\) being the total observation time encompassing \(M\) cycles. The numerical simulation in Fig.\,\ref{Fig.stability}(a) presents normalized \(c_1:c_2:c_3 \approx 1.00:0.09:0.01\), with other harmonic components being neglected. The adoption of multi-harmonic measurements in the HSR regime, as opposed to relying solely on the fundamental frequency or a single harmonic component, shows a potential to differentiate signal characteristics, thereby enhancing the signal-to-noise ratio (SNR)\,\cite{wu2022experimentalPRL}.

\section*{Acknowledgement}

We thank Lei Cong for helpful discussions. K.W. was funded by the Innovation Program for Quantum Science and Technology under Grant 2021ZD0300401,  by the National Science Foundation of China (NSFC) under Grants No. 62203030 and 61925301 for Distinguished Young Scholars, by the Fundamental Research Founds for the Central Universities. D.B. was funded by the DFG Project ID 390831469: EXC 2118 (PRISMA+ Cluster of Excellence), by the COST Action within the project COSMIC WISPers (Grant No. CA21106), and by the QuantERA project LEMAQUME (DFG Project No. 500314265). W.J. was funded by the Peking University Startup Fund. 

\bibliographystyle{utphys}
%\bibliography{mainrefs}

\begin{thebibliography}{10}

\bibitem{wu1957}
C.~S. Wu, E.~Ambler, R.~W. Hayward, D.~D. Hoppes, and R.~P. Hudson,
  ``Experimental test of parity conservation in beta decay,''
  \href{http://dx.doi.org/10.1103/PhysRev.105.1413}{{\em Phys. Rev.} {\bfseries
  105} (Feb, 1957) 1413--1415}.
  \url{https://link.aps.org/doi/10.1103/PhysRev.105.1413}.

\bibitem{barkov1978}
L.~M. Barkov and M.~S. Zolotorev, ``Pis’ma zh. eksp. teor. fiz. 27, 379 [jetp
  lett. 27, 357 (1978)],'' {\em Pis’ma Zhurnal Éksperimental’noĭ i
  Teoreticheskoĭ Fiziki} {\bfseries 27} (1978) 379. JETP Lett. 27, 357 (1978).

\bibitem{conti1979}
R.~Conti, P.~Bucksbaum, S.~Chu, E.~Commins, and L.~Hunter, ``Preliminary
  observation of parity nonconservation in atomic thallium,''
  \href{http://dx.doi.org/10.1103/PhysRevLett.42.343}{{\em Phys. Rev. Lett.}
  {\bfseries 42} (Feb, 1979) 343--346}.
  \url{https://link.aps.org/doi/10.1103/PhysRevLett.42.343}.

\bibitem{wood1997}
C.~S. Wood, S.~C. Bennett, D.~Cho, B.~P. Masterson, J.~L. Roberts, C.~E.
  Tanner, and C.~E. Wieman, ``Measurement of parity nonconservation and an
  anapole moment in cesium,''
  \href{http://dx.doi.org/10.1126/science.275.5307.1759}{{\em Science}
  {\bfseries 275} no.~5307, (1997) 1759--1763}.

\bibitem{bouchiat2011atomic}
M.-A. Bouchiat, ``Atomic parity violation. early days, present results,
  prospects,'' {\em arXiv preprint arXiv:1111.2172} (2011) .

\bibitem{Gwinner2022}
G.~Gwinner and L.~A. Orozco, ``Studies of the weak interaction in atomic
  systems: towards measurements of atomic parity non-conservation in
  francium,'' \href{http://dx.doi.org/10.1088/2058-9565/ac4424}{{\em Quantum
  Science and Technology} {\bfseries 7} no.~2, (Jan, 2022) 024001}.
  \url{https://dx.doi.org/10.1088/2058-9565/ac4424}.

\bibitem{tsigutkin2009}
K.~Tsigutkin, D.~Dounas-Frazer, A.~Family, J.~E. Stalnaker, V.~V. Yashchuk, and
  D.~Budker, ``Observation of a large atomic parity violation effect in
  ytterbium,'' \href{http://dx.doi.org/10.1103/PhysRevLett.103.071601}{{\em
  Physical Review Letters} {\bfseries 103} no.~7, (2009) 071601}.

\bibitem{antypas_isotopic_2019}
D.~Antypas, A.~Fabricant, J.~E. Stalnaker, K.~Tsigutkin, V.~V. Flambaum, and
  D.~Budker, ``Isotopic variation of parity violation in atomic ytterbium,''
  \href{http://arxiv.org/abs/1804.05747}{{\ttfamily 1804.05747}}.
  \url{http://dx.doi.org/10.1038/s41567-018-0312-8}. Publisher: Springer {US}.

\bibitem{safronova2018search}
M.~Safronova, D.~Budker, D.~DeMille, D.~F.~J. Kimball, A.~Derevianko, and C.~W.
  Clark, ``Search for new physics with atoms and molecules,''
  \href{http://dx.doi.org/10.1103/RevModPhys.90.025008}{{\em Rev. Mod. Phys.}
  {\bfseries 90} no.~2, (2018) 025008}.

\bibitem{dobrescu2006spin}
B.~A. Dobrescu and I.~Mocioiu, ``Spin-dependent macroscopic forces from new
  particle exchange,''
  \href{http://dx.doi.org/10.1088/1126-6708/2006/11/005}{{\em J. High Energy
  Phys.} {\bfseries 2006} no.~11, (2006) 005}.

\bibitem{moody1984new}
J.~Moody and F.~Wilczek, ``New macroscopic forces?,''
  \href{http://dx.doi.org/10.1103/PhysRevD.30.130}{{\em Phys. Rev. D}
  {\bfseries 30} (1984) 130}.

\bibitem{fadeev2019revisiting}
P.~Fadeev, Y.~V. Stadnik, F.~Ficek, M.~G. Kozlov, V.~V. Flambaum, and
  D.~Budker, ``Revisiting spin-dependent forces mediated by new bosons:
  potentials in the coordinate-space representation for macroscopic-and
  atomic-scale experiments,''
  \href{http://dx.doi.org/10.1103/PhysRevA.99.022113}{{\em Phys. Rev. A}
  {\bfseries 99} no.~2, (2019) 022113}.

\bibitem{cong2024spin}
L.~Cong, W.~Ji, P.~Fadeev, F.~Ficek, M.~Jiang, V.~V. Flambaum, H.~Guan,
  D.~F.~J. Kimball, M.~G. Kozlov, Y.~V. Stadnik, {\em et~al.}, ``Spin-dependent
  exotic interactions,''
  \href{http://dx.doi.org/10.48550/arXiv.2408.15691}{{\em Rev. Mod. Phys.}
  ((2025)) }.

\bibitem{wei2023ultrasensitive}
K.~Wei, T.~Zhao, X.~Fang, Z.~Xu, C.~Liu, Q.~Cao, A.~Wickenbrock, Y.~Hu, W.~Ji,
  J.~Fang, {\em et~al.}, ``Ultrasensitive atomic comagnetometer with enhanced
  nuclear spin coherence,''
  \href{http://dx.doi.org/10.1103/PhysRevLett.130.063201}{{\em Phys. Rev.
  Lett.} {\bfseries 130} no.~6, (2023) 063201}.

\bibitem{heng2023ultrasensitive}
X.~Heng, K.~Wei, T.~Zhao, Z.~Xu, Q.~Cao, X.~Huang, Y.~Zhai, M.~Ye, and W.~Quan,
  ``Ultrasensitive optical rotation detection with closed-loop suppression of
  spin polarization error,'' {\em IEEE Transactions on Instrumentation and
  Measurement} {\bfseries 72} (2023) 1--12.

\bibitem{xu2023darkmatter}
K.~Wei, Z.~Xu, Y.~He, X.~Ma, X.~Heng, X.~Huang, W.~Quan, W.~Ji, J.~Liu, X.-P.
  Wang, {\em et~al.}, ``Dark matter search with a resonantly-coupled hybrid
  spin system,'' {\em Reports on Progress in Physics} (2025) .

\bibitem{su2021search}
H.~Su, Y.~Wang, M.~Jiang, W.~Ji, P.~Fadeev, D.~Hu, X.~Peng, and D.~Budker,
  ``Search for exotic spin-dependent interactions with a spin-based
  amplifier,'' \href{http://dx.doi.org/10.1126/sciadv.abi9535}{{\em Sci. Adv.}
  {\bfseries 7} no.~47, (2021) eabi9535}.

\bibitem{yan2015searching}
H.~Yan, G.~Sun, S.~Peng, Y.~Zhang, C.~Fu, H.~Guo, and B.~Liu, ``Searching for
  new spin-and velocity-dependent interactions by spin relaxation of polarized
  he 3 gas,'' {\em Physical Review Letters} {\bfseries 115} no.~18, (2015)
  182001.

\bibitem{lee2018improved}
J.~Lee, A.~Almasi, and M.~Romalis, ``Improved limits on spin-mass
  interactions,'' {\em Physical review letters} {\bfseries 120} no.~16, (2018)
  161801.

\bibitem{liang2023new}
H.~Liang, M.~Jiao, Y.~Huang, P.~Yu, X.~Ye, Y.~Wang, Y.~Xie, Y.-F. Cai, X.~Rong,
  and J.~Du, ``New constraints on exotic spin-dependent interactions with an
  ensemble-nv-diamond magnetometer,'' {\em National Science Review} {\bfseries
  10} no.~7, (2023) nwac262.

\bibitem{crescini2022search}
N.~Crescini, G.~Carugno, P.~Falferi, A.~Ortolan, G.~Ruoso, and C.~Speake,
  ``Search of spin-dependent fifth forces with precision magnetometry,'' {\em
  Physical Review D} {\bfseries 105} no.~2, (2022) 022007.

\bibitem{xu2022critical}
Z.~Xu, K.~Wei, X.~Heng, X.~Huang, and Y.~Zhai, ``Critical dynamics of strongly
  interacting ensembles in spin-exchange-relaxation-free comagnetometers,''
  \href{http://dx.doi.org/10.1103/PhysRevApplied.18.044049}{{\em Phys. Rev.
  Applied} {\bfseries 18} no.~4, (2022) 044049}.

\bibitem{padniuk2024universal}
M.~Padniuk, E.~Klinger, G.~{\L}ukasiewicz, D.~Gavilan-Martin, T.~Liu,
  S.~Pustelny, D.~F. Jackson~Kimball, D.~Budker, and A.~Wickenbrock,
  ``Universal determination of comagnetometer response to spin couplings,''
  {\em Physical Review Research} {\bfseries 6} no.~1, (2024) 013339.

\bibitem{wei2022constraints}
K.~Wei, W.~Ji, C.~Fu, A.~Wickenbrock, V.~V. Flambaum, J.~Fang, and D.~Budker,
  ``Constraints on exotic spin-velocity-dependent interactions,''
  \href{http://dx.doi.org/10.1038/s41467-022-34924-z}{{\em Nat. Commun.}
  {\bfseries 13} no.~1, (2022) 7387}.

\bibitem{almasi2020new}
A.~Almasi, J.~Lee, H.~Winarto, M.~Smiciklas, and M.~V. Romalis, ``New limits on
  anomalous spin-spin interactions,''
  \href{http://dx.doi.org/10.1103/PhysRevLett.125.201802}{{\em Phys. Rev.
  Lett.} {\bfseries 125} no.~20, (2020) 201802}.

\bibitem{ji2018new}
W.~Ji, Y.~Chen, C.~Fu, M.~Ding, J.~Fang, Z.~Xiao, K.~Wei, and H.~Yan, ``New
  experimental limits on exotic spin-spin-velocity-dependent interactions by
  using smco 5 spin sources,'' {\em Physical review letters} {\bfseries 121}
  no.~26, (2018) 261803.

\bibitem{xu2025new}
Z.~Xu, X.~Heng, G.~Tian, D.~Gong, L.~Cong, W.~Ji, D.~Budker, and K.~Wei, ``New
  constraints on axion mediated dipole-dipole interactions,'' {\em arXiv
  preprint arXiv:2501.07865} (2025) .

\bibitem{ji2023constraints}
W.~Ji, W.~Li, P.~Fadeev, F.~Ficek, J.~Qin, K.~Wei, Y.-C. Liu, and D.~Budker,
  ``Constraints on spin-spin velocity-dependent interactions,''
  \href{http://dx.doi.org/10.1103/PhysRevLett.130.133202}{{\em Phys. Rev.
  Lett.} {\bfseries 130} no.~13, (2023) 133202}.

\bibitem{liu2023modeling}
C.~Liu, Z.~Xu, K.~Wei, D.~Gong, X.~Heng, X.~Huang, W.~Quan, and Y.~Zhai,
  ``Modeling and suppression of atomic comagnetometer’s response to
  micro-vibration,'' {\em Sensors and Actuators A: Physical} {\bfseries 359}
  (2023) 114503.

\bibitem{tian2025active}
G.~Tian, X.~Gong, M.~Xia, D.~Gong, K.~Qi, C.~Liu, R.~Wang, and K.~Wei, ``Active
  control of low-frequency vibrations with parameter self-optimization for
  quantum sensing,'' {\em Measurement} (2025) 117195.

\bibitem{wu2022experimentalPRL}
K.~Wu, S.~Chen, G.~Sun, S.~Peng, M.~Peng, and H.~Yan, ``Experimental limits on
  exotic spin and velocity dependent interactions using rotationally modulated
  source masses and an atomic-magnetometer array,'' {\em Physical Review
  Letters} {\bfseries 129} no.~5, (2022) 051802.

\bibitem{kim2019experimentalNC}
Y.~J. Kim, P.-H. Chu, I.~Savukov, and S.~Newman, ``Experimental limit on an
  exotic parity-odd spin-and velocity-dependent interaction using an optically
  polarized vapor,'' {\em Nature Communications} {\bfseries 10} no.~1, (2019)
  2245.

\bibitem{yan2013new}
H.~Yan and W.~Snow, ``New limit on possible long-range parity-odd interactions
  of the neutron from neutron-spin rotation in liquid \text{$^4$He},''
  \href{http://dx.doi.org/10.1103/PhysRevLett.110.082003}{{\em Phys. Rev.
  Lett.} {\bfseries 110} no.~8, (2013) 082003}.

\bibitem{clayburn2023using}
N.~B. Clayburn and L.~R. Hunter, ``Using earth to search for long-range
  spin-velocity interactions,'' {\em Physical Review D} {\bfseries 108} no.~5,
  (2023) L051701.

\bibitem{aasi2015advanced}
J.~Aasi, B.~Abbott, R.~Abbott, T.~Abbott, M.~Abernathy, K.~Ackley, C.~Adams,
  T.~Adams, P.~Addesso, R.~Adhikari, {\em et~al.}, ``Advanced ligo,'' {\em
  Classical and quantum gravity} {\bfseries 32} no.~7, (2015) 074001.

\bibitem{duan2025concurrent}
J.~Duan, Z.~Hu, X.~Lu, L.~Xiao, S.~Jia, K.~M{\o}lmer, and Y.~Xiao, ``Concurrent
  spin squeezing and field tracking with machine learning,'' {\em Nature
  Physics} (2025) 1--7.

\bibitem{troullinou2021squeezed}
C.~Troullinou, R.~Jim{\'e}nez-Mart{\'\i}nez, J.~Kong, V.~Lucivero, and
  M.~Mitchell, ``Squeezed-light enhancement and backaction evasion in a high
  sensitivity optically pumped magnetometer,'' {\em Physical Review Letters}
  {\bfseries 127} no.~19, (2021) 193601.

\bibitem{zel1959parity}
Y.~B. Zel’dovich, ``Parity nonconservation in the first order in the
  weak-interaction constant in electron scattering and other effects,'' {\em
  Journal of Experimental and Theoretical Physics (JETP)} {\bfseries 36} (1959)
  964--966.

\bibitem{xu2024constraining}
Z.~Xu, X.~Ma, K.~Wei, Y.~He, X.~Heng, X.~Huang, T.~Ai, J.~Liao, W.~Ji, J.~Liu,
  {\em et~al.}, ``Constraining ultralight dark matter through an accelerated
  resonant search,'' \href{http://dx.doi.org/10.1038/s42005-024-01713-7}{{\em
  Commun. Phys.} {\bfseries 7} no.~1, (2024) 226}.

\bibitem{wu2022searching}
K.~Wu, S.~Chen, J.~Gong, M.~Peng, and H.~Yan, ``Searching for exotic
  spin-dependent interactions using rotationally modulated source masses and an
  atomic magnetometer array,'' {\em Physical Review D} {\bfseries 105} no.~5,
  (2022) 055020.

\end{thebibliography}
\providecommand{\href}[2]{#2}\begingroup\raggedright\endgroup

\end{document}